\documentclass[aps,twocolumn,superscriptaddress,nofootinbib,longbibliography,notitlepage,floatfix]{revtex4-2}
\usepackage{graphicx}
\usepackage{amssymb}
\usepackage{hyperref}
\usepackage{amsmath}
\usepackage{array}
\usepackage{booktabs}
\usepackage{multirow}
\usepackage{amsmath}
\usepackage{amssymb}
\usepackage{graphicx}
\usepackage{esint}

\usepackage{amsfonts}
\usepackage{subfigure}
\usepackage{dsfont}
\usepackage{times}
\usepackage{dcolumn}
\usepackage{esint}
\usepackage{mathrsfs}
\usepackage{bm}
\usepackage{multirow}
\usepackage{color}
\usepackage{datetime}
\usepackage{braket}

\usepackage{graphicx}
\usepackage{dcolumn}
\usepackage{bm}
\usepackage{hyperref}
\usepackage{xcolor}
\usepackage{physics}
\usepackage{braket}

\begin{document}

 \title{Entanglement Fractalization}

\author{Yao Zhou}
\affiliation{Guangdong Provincial Key Laboratory of Magnetoelectric Physics and Devices, State Key Laboratory of Optoelectronic Materials and Technologies, and School of Physics, Sun Yat-sen University, Guangzhou, 510275, China}

\author{Peng Ye}\email{yepeng5@mail.sysu.edu.cn}
\affiliation{Guangdong Provincial Key Laboratory of Magnetoelectric Physics and Devices, State Key Laboratory of Optoelectronic Materials and Technologies, and School of Physics, Sun Yat-sen University, Guangzhou, 510275, China}

\date{\today}

\begin{abstract}
We numerically explore the interplay of fractal geometry and quantum entanglement by analyzing the von Neumann entropy (known as entanglement entropy) and the entanglement contour in the scaling limit. 
Adopting quadratic fermionic models on Sierpinski carpet, we uncover intriguing findings.
For gapless ground states exhibiting a finite density of states at the chemical potential, we reveal a super-area law characterized by the presence of a logarithmic correction for area law in the scaling of entanglement entropy. This extends the well-established super-area law observed on translationally invariant Euclidean lattices where the Gioev-Klich-Widom conjecture regarding the asymptotic behavior of Toeplitz matrices holds significant influence.  Furthermore, different from the fractal structure of the lattice, we observe the emergence of a novel self-similar and universal pattern termed an ``entanglement fractal'' in the entanglement contour data as we approach the scaling limit. 
Remarkably, this pattern bears resemblance to intricate Chinese paper-cutting designs. We provide  general rules to artificially generate this fractal, offering insights into the universal scaling of entanglement entropy. Meanwhile, as the direct consequence of the entanglement fractal and beyond a single scaling behavior of entanglement contour in translation-invariant systems, we identify two distinct scaling behaviors in the entanglement contour of fractal systems.
For gapped ground states, we observe that the entanglement entropy adheres to a generalized area law, with its dependence on the Hausdorff dimension of the boundary between complementary subsystems.
\end{abstract}

\maketitle

\section{Introduction}
Entanglement provides a quantum-informative perspective for understanding non-local correlations in many-body quantum systems~\cite{amicoEntanglement2008,eisertColloquium2010,laflorencieQuantum2016}. Quantitatively, entanglement entropy (EE), or von Neumann entropy, stands out as one of the most useful quantities for characterizing the strength of quantum entanglement. For quantum many-body systems defined on Euclidean lattice with  dimension $d_s$, it is well-known that the EE of gapped ground states follows a universal scaling feature called the area law\footnote{Here, the term ``area'' specifically refers to the boundary area between the two complementary subsystems, denoted as $A$ and $B$.}, given by $S\sim L_A^{d_s-1}$, where $L_A$ is the linear size of the boundary between the two complementary subsystems $A$ and $B$~\cite{eisertColloquium2010,laflorencieQuantum2016,hastingsArea2007}. Interestingly, gapped phases hosting anyons exhibit a universal constant entropy known as topological entanglement entropy. This appears as the subleading term in EE, quantitatively fixed by the total quantum dimension of all anyons~\cite{kitaevTopological2006,levinDetecting2006}.

On the other hand, for gapless free-fermion systems defined on translationally invariant Euclidean lattice with dimension $d_s$, when  the Fermi surface is of codimension-$1$, the EE scales as a ``super area law'', given by $S\sim L_A^{d_s-1}\log L_A$. This scaling is derived by applying  Gioev-Klich-Widom (Widom) conjecture of asymptotic behavior of Toeplitz matrices and numerically confirmed~\cite{gioevEntanglement2006,wolfViolation2006,liScaling2006,barthelEntanglement2006,swingleEntanglement2010,dingEntanglement2012,laflorencieQuantum2016,linMeasuring2024}, where translation symmetry plays a vital role. The scaling also demonstrates how logarithmic correction is induced by the infinite number of gapless modes on the Fermi surface~\cite{swingleEntanglement2010}. However, the situation differs for Fermi surfaces of higher codimension. For example, in 2D systems with Fermi points, such as graphene with Dirac points, the EE still follows the area law. This implies that the vanishing density-of-state of Fermi points cannot provide sufficient long-range quantum correlation to enhance quantum entanglement between the two complementary subsystems.

All the aforementioned investigations into fermion system entanglement are conducted under the assumption of a translationally invariant lattice. However, fermions may also move on a fractal lattice~\cite{mandelbrot1983fractal} embedded in some Euclidean space, resulting in a scenario where fermions experience fractional dimensionality and translation symmetry is absent. As a type of very exotic geometric patterns, fractals, common in nature, have attracted academic interest since the last century, with statistical models and critical phenomena on fractals gaining substantial research attention~\cite{gefenCritical1980,kaufmanExactly1981,gefenGeometric1983,gefenPhase1984,tasakiCritical1987,komaClassical1995,yoshidaQuantum2014}. Ongoing efforts continue to explore various physical phenomena~\cite{shangAssembling2015a,liConstruction2017,kempkesDesign2019a,moOnSurface2019,liuSierpi2021,vanveenOptical2017,kempkesDesign2019a,paiTopological2019,iliasovHall2020,biesenthalFractal2022,yangElectronic2022,mannaHigherorder2022,liHigherorder2022a,zhuTopological2022,ivakiTopological2022,duaQuantum2023,jhaProperties2023a,zhanghigh2024}, including topological effects, optical and transport properties, and the practical implementation of fractal topological quantum memory. Among these research works, the numerical simulation is the predominant method due to the complexity of fractals. These findings are substantially influenced by the unique features of fractals: fractional dimensions affecting microscopic degrees of freedom and self-similarity with broken translation symmetry. Therefore, considering that the aforementioned super-area law depends on translation symmetry, enabling the application of the mathematical Widom conjecture of Toeplitz matrices, it is natural to explore the scaling law of the EE in many-body systems defined on fractal lattices. These systems lack translation symmetry and have non-integer dimensions, where the Widom conjecture  is no longer applicable.  This line of thinking, exploring the interplay of quantum entanglement and fractal geometry, serves as the motivation for the present work. Notably, the previous works are focused on studying the EE of finitely ramified fractal by using the heat-kernel method~\cite{farajiastanehEntanglement2016,akalEntanglement2018}, where this method can not be applied to the infinitely ramified fractal. Ultimately, we aim to gain a deeper understanding of how quantum entanglement is influenced by infinitely ramified fractals where macroscopic number of fermionic quantum particles move.

In this paper, we investigate the entanglement properties of many-body systems on the Sierpinski carpet, a representative infinitely ramified fractal, using quadratic fermionic models. Our analysis focuses on two key aspects: the entanglement entropy (EE) and the entanglement contour (EC). Mathematically, the EE can be decomposed into site-dependent contributions denoted as $s(i)$: $S_A = \sum_{i \in A} s(i)$, as introduced for Gaussian states~\cite{chenEntanglement2014}. These site-dependent terms, $s(i)$, collectively form the EC, which provides a real-space ``tomography'' of quantum entanglement, offering insight into the fine structure of the real-space distribution of EE~\cite{moUnraveling2023}. 
The EC has been employed to reveal the fine structure of entanglement in various fields, including conformal field theory, geometric constructions in AdS/CFT correspondence, and disordered systems~\cite{Tonnientanglement2018,Wenfine2018,royEntanglement2018}. This fine structure, $s(i)$, offers a direct visualization of the scaling behavior of EE. For instance, in gapped systems, $s(i)$ decays exponentially with the distance from the boundary, implying that only the degrees of freedom near the boundary contribute significantly to the EE. As a result, the EE in such systems should satisfy the area law~\cite{chenEntanglement2014}. On the other hand, in disordered systems, the EC provides a framework to explain the violation of the area law in the delocalized phase~\cite{royEntanglement2018}.

\begin{table}
\caption{\label{tab1} Comparison of the density of states (DOS) , scaling of entanglement entropy $S_{A}$, and the scaling and distribution of the entanglement contour $s_{i}$ with respect to the distance to the boundary between subsystems on a normal lattice (NL) with translation invariance and the Sierpinski carpet (SC) with self-similarity.
}
\begin{ruledtabular}
\begin{tabular}{cccccc}
\multirow{2}{*}{Lattice} & \multirow{2}{*}{Phase} & \multirow{2}{*}{DOS} & \multirow{2}{*}{$S_{A}$}  & \multicolumn{2}{c}{$s_{i}$}    \\ 
& &  & & Scaling & Distribution  \\ \hline
\multirow{2}{*}{NL}& Gapped & $0$ & $\sim L_{A}^{d_{s}-1}$ & Exp. decay & Uniform\\
& Gapless & $>0$ & $\sim L_{A}^{d_{s}-1}\log L_{A}$ & Power-law decay &  Uniform \\ \hline
\multirow{2}{*}{SC}& Gapped & $0$ & $\sim L_{A}^{d_{\rm bf}}$ & Exp. decay  & Uniform \\
& Gapless & $>0$ & $\sim L^{d_{s}-1}_{A} \log L_{A}$ & Power-law decay &  Fractal \\
\end{tabular}
\end{ruledtabular}
\end{table}

Our initial examination considers a gapless system embedded in 2D space with a finite density-of-state (DOS) at the chemical potential within the fractal lattice. To assess entanglement between the two complementary subsystems (labeled as $A$ and $B$) of a pure state, we employ various bi-partition schemes, which may break self-similar and even allow for the fractal structure of the boundary between $A$ and $B$. For generic free-fermion systems on fractal lattices with a finite DOS at the chemical potential, we observe a super-area scaling of EE: $S_{A}\sim L_{A}^{d_{s}-1} \log L_{A}$, where $d_s=2$ denotes the dimension of the embedding space in our numerical setting. This scaling result found in the numerical computation extends generalizes the applicability of the super-area law from translationally invariant Euclidean lattices to fractal lattices, as shown in Tab.~\ref{tab1}.

\begin{figure}
\centering
\includegraphics[width=8.6cm]{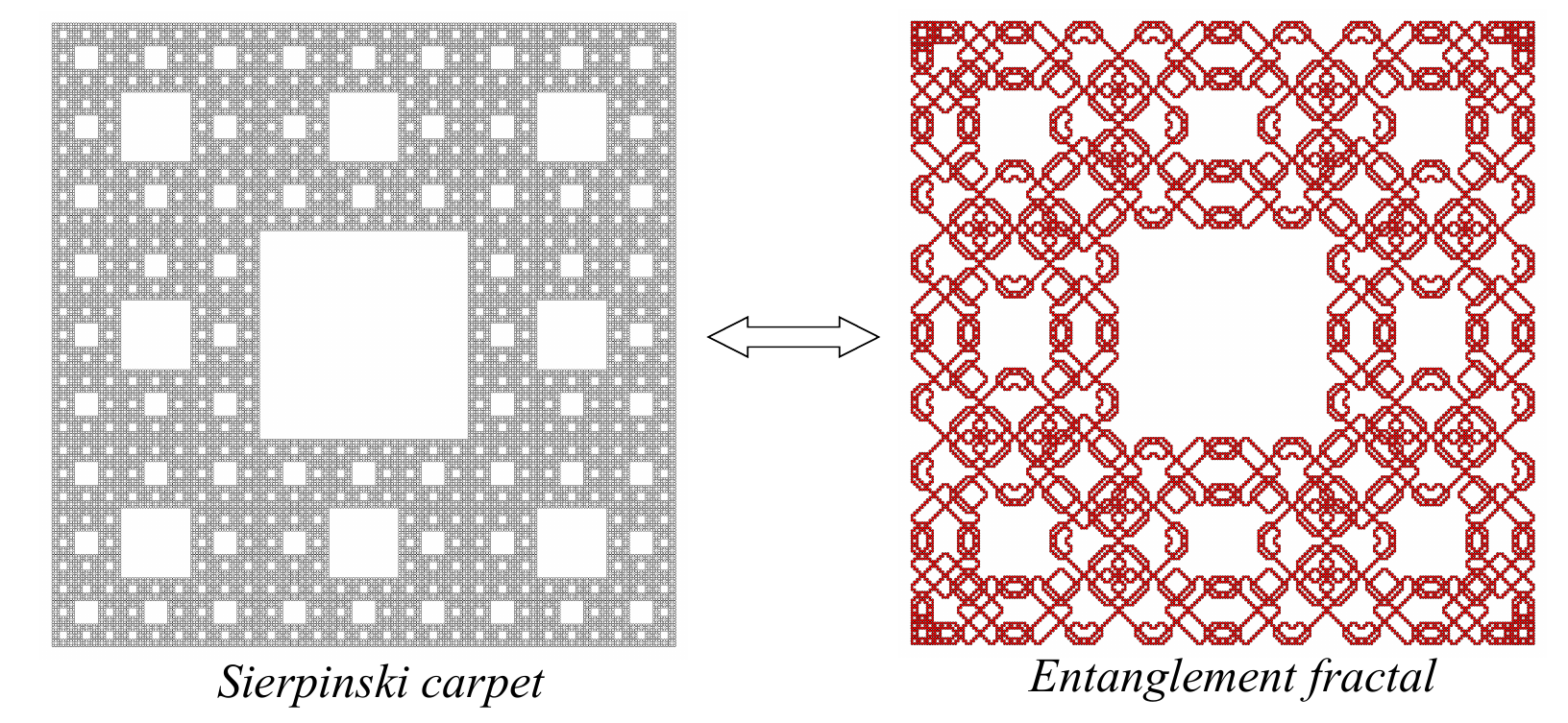}
\caption{Pictorial illustration of fractal lattice and  entanglement fractal which resembles Chinese papercutting. See main texts for details.
}\label{fig_scdos}

\end{figure}

Next, we focus on the EC data and reveal a fascinating discovery: the emergence of a self-similar, universally present fractal pattern distinct from the Sierpinski carpet's fractal structure, which we term the ``entanglement fractal'' (EF) pattern (see Fig.~\ref{fig_scdos}). In contrast to the uniform distribution of EC observed in translation-invariant systems, as shown in Tab.~\ref{tab1}, the EF pattern exhibits intricate features reminiscent of the elaborate designs found in Chinese paper-cutting artistry. Remarkably, despite being identified through numerical computations, the EF pattern can be generated using a set of general rules. Furthermore, we observe that the entanglement contour exhibits two distinct scaling behaviors, which is the direct consequence of the EF and not previously seen in translation-invariant systems. This observation indicates that only a portion of the physical degrees of freedom significantly contributes to the EE, highlighting a complex relationship between entanglement and system geometry.

For gapped systems, our numerical analysis proposes a universal scaling of EE as $S_{A} \sim L_{A}^{d_{\rm bf}}$, where $d_{\rm bf}$ denotes the Hausdorff dimension of $A$'s boundary, thereby extending beyond the conventional area law. Furthermore, in gapped systems, the EC predominantly resides at $A$'s boundary and exhibits a uniformly exponential decay into the bulk in Tab.~\ref{tab1}.

\section{Definition of entanglement entropy and entanglement contour}\label{section_defini}
In the beginning, we give a brief discussion for the definition of the EE and the EC. 
Consider a many-body ground state $ \ket{G} $ with density matrix $ \rho=\ket{G}\bra{G} $. When partitioning the system into two subsystems $ A $ and $ B $, the reduced density matrix can be obtained by tracing over subsystem $ B $:
\begin{align}
\rho_{A} = \text{Tr}_{B}\ket{G}\bra{G} = \frac{1}{\mathcal{N}}\exp(-\mathcal{H}^{E})\,,
\end{align}
where $\mathcal{N}$ is a normalization constant. In the free-fermion limit, owing to the quadratic form of the entanglement Hamiltonian~\cite{peschelCalculation2003,klichLower2006}:
\begin{align}
\mathcal{H}^{E} = \sum_{i,j}c^{\dagger}_{i}h^{E}_{i,j}c_{j} 
\end{align}
with $ i,j \in A $, where $ c^{\dagger}_{i}(c_{i})$ represents a fermionic creation (annihilation) operator at the $i$-th lattice site. Consequently, entanglement information can be extracted from the entanglement Hamiltonian matrix $h^{E}$. Additionally, as discussed in Ref.~\cite{peschelCalculation2003,klichLower2006}, the spectrum $ \{\varepsilon_{\iota}\}$ of $h^{E}$ and the spectrum $\{\xi_{\iota}\}$ of the correlation matrix $C^{A}(i,j) = \bra{G}c^{\dagger}_{i}c_{j}\ket{G}$ with $i,j$ in subsystem $A$ have a one-to-one correspondence with the same eigenvector, expressed as:
\begin{align}
\varepsilon_{\iota}=\log[(\xi_{\iota})^{-1}-1]\,.
\end{align}
For convenience in free-fermion systems, the spectrum $\{\xi_{\iota}\}$ is typically adopted as the entanglement spectrum, and the entanglement entropy (EE) is calculated using the equation:
\begin{align}
S_{A} = \text{Tr}[f(C^{A})]= -\sum_{\iota}[\xi_{\iota}\log \xi_{\iota}+(1-\xi_{\iota})\log(1-\xi_{\iota})]\,. 
\end{align}
Utilizing the eigenvalues and eigenvectors of the correlation matrix $C^{A}$, the entanglement contour (EC) in free-fermion systems is defined as:
\begin{equation}
s(i) = \sum_{\iota}p_{i}(\iota)S_{\iota}\,,
\label{eq_ec_def}
\end{equation}
where $S_{\iota} =-[\xi_{\iota}\log \xi_{\iota}+(1-\xi_{\iota})\log(1-\xi_{\iota})]$. $p_{i}(\iota) = |\braket{i}{\iota}|^{2}$ represents the probability of the eigenvector $\ket{\iota}$ of the correlation matrix at site $i$ in $A$, and $\sum_{i}p_{i}(\iota) =1$. These eigenvectors, often referred to as ``Schmidt vectors'' or ``entanglement wavefunctions'', encapsulate information regarding bulk-boundary correspondence in topological gapless systems~\cite{zhouEntanglement2023}.

\section{EE of gapless ground states}
At the outset, prior to diving into the examination of entanglement features within fractal systems, it is imperative to address the process of constructing a lattice system exhibiting a fractal structure. Initially, we introduce two variable elements crucial to the fractal lattice—namely, an initial unit cell and a fractal iteration method. These components uniquely determine a specific type of fractal lattice. Employing the iteration method iteratively $n$ times on the unit cell with lattice sites yields the $n$th-order approximation of the fractal. For instance, the $5$th-order approximation $SC(5,1)$ of the Sierpinski carpet~\cite{mandelbrot1983fractal}, depicted in the upper part of Fig.~\ref{fig_scdos} (refer to \textbf{Appendix}~\ref{app_fractal_a} and Fig.~\ref{fig_frac_iter} for detailed information; note that ``$1$'' in $SC(5,1)$ represents the width of a unit cell).

\begin{figure}[htbp]
\centering
\includegraphics[width=8.6cm]{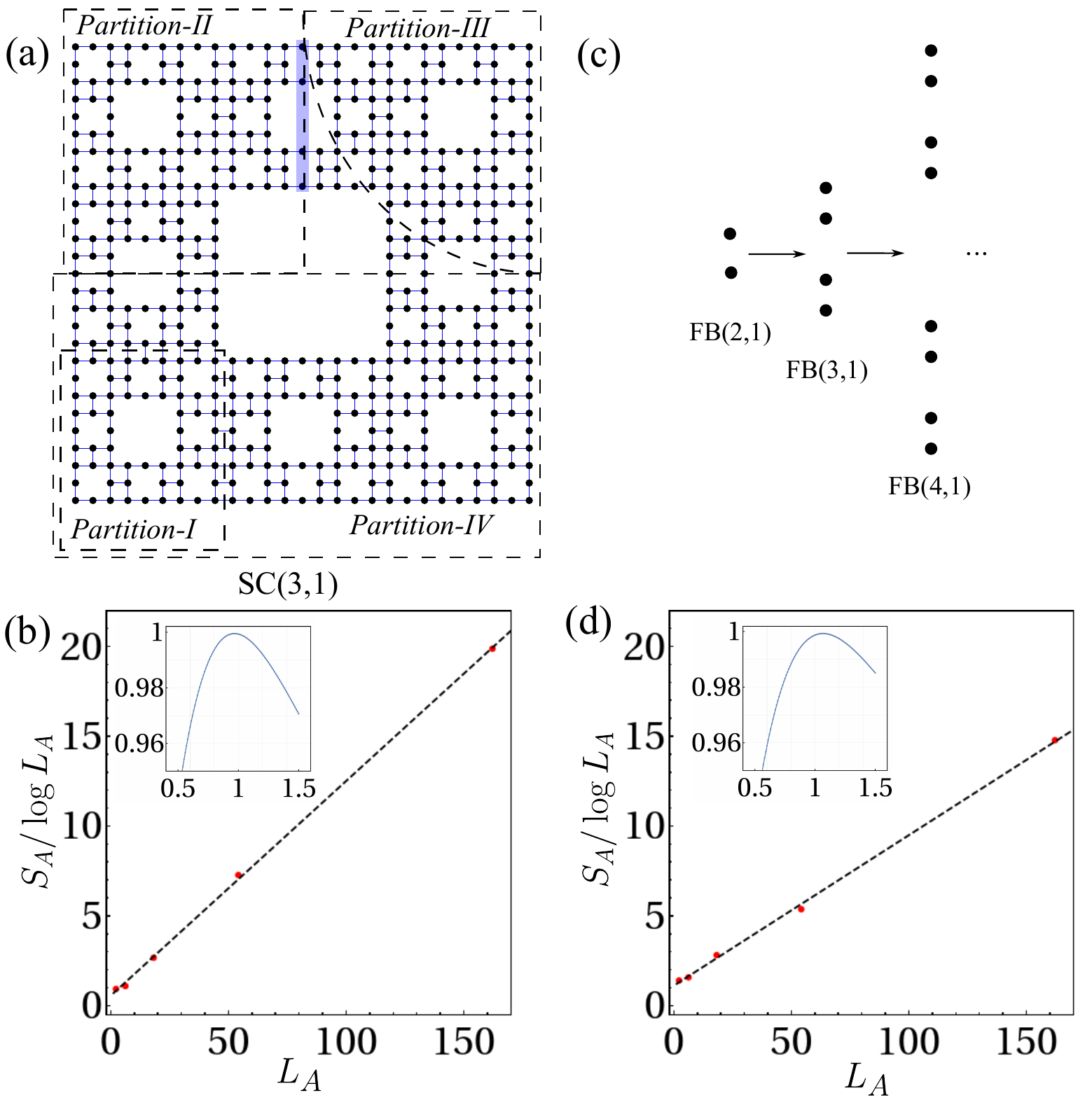}
\caption{
(a) Four kinds of partitions on the $3$rd-order approximation $SC(3,1)$ of Sierpinski carpet as an example. Partition-I has a normal boundary and Partition-II has a fractal boundary in the blue area. (c) The fractal structure of the boundary for different $n$. (b) and (d) The EE of the model on the $n$th-order approximation $SC(n,1)$ of Sierpinski carpet, using Partition-I and II, respectively, where the dashed line is the linear fit of EE data. The insets show the coefficient of determination $R^{2}$ as a function of $\alpha$ to measure the goodness of fit. Here $t=1$ and $\mu=0$.}\label{fig_part_frac}
\end{figure}

Subsequently, we investigate the scaling of EE for the gapless system defined on the fractal lattice $SC(n,1)$ embedded in a $2$-dimensional spatial space, as illustrated in Fig.~\ref{fig_part_frac}(a). We consider a spinless tight-binding model described by the Hamiltonian:
\begin{equation}
H_{1} = -t \sum_{<ij>} c^{\dagger}_{i} c_{j} - \mu \sum_{i}c^{\dagger}_{i} c_{i}\,.
\end{equation}
Here, $c^{\dagger}_{i}(c_{i})$ represents a fermionic creation (annihilation) operator at the $i$-th lattice site, $<ij>$ denotes nearest-neighbor sites, and $\mu$ is the chemical potential. The model $H_{1}$ exhibits symmetries due to the properties of the fractal lattice $SC(n,1)$, such as a four-fold rotation symmetry. To ascertain the bulk gap, it is necessary to compute the DOS of the model $H_{1}$ on the fractal lattice $SC(n,1)$. Given the exponential increase in the number of lattice sites with each iteration of the fractal lattice~\cite{mandelbrot1983fractal}, the numerical exploration of entanglement presents considerable challenges. Specifically, numerical calculations in entanglement for translation-invariant systems hinge on the diagonalization of a non-sparse matrix at a large size limit. The challenge is further amplified when translation symmetry is absent, necessitating a computationally intensive numerical study. In our case, through an analysis of the scaling of the energy gap and the DOS, we demonstrate that the model $H_{1}$ on the Sierpinski carpet exhibits a gapless ground state with a finite DOS at the chemical potential (see \textbf{Appendix}~\ref{app_scaling_b} and Fig.~\ref{fig_gap_gapless} for technical details), indicative of a metalic ground state on the fractal lattice.

To progress further, we initially establish various bi-partition schemes in order to compute entanglement quantities. In Fig.~\ref{fig_part_frac} (a), we present the $3$rd-order approximation $SC(3,1)$ of the Sierpinski carpet as an illustrative example. Four distinct partitioning methods (labeled I-IV) for dividing the original fractal lattice into subsystems $A$ and $B$ are demonstrated, where the   area enclosed by dashed lines is designated as the subsystem $A$ for each partition scheme. Specifically, for Partition-I, the subsystem $A$ geometrically corresponds to the $2$nd-order approximation $SC(2,1)$ that preserves all spatial symmetry and self-similar of the original fractal lattice. In a general sense, for the $n$th-order approximation $SC(n,1)$, $A$ represents the $(n-1)$th-order approximation $SC(n-1,1)$ with linear length $L_{A}= 3^{n-1} l$ and the number of boundary sites $N_{b A}=3^{n-1}$, where the lattice constant $l=1$. By noting $L_{A}=N_{bA}$, the boundary of $A$ using Partition-I in Fig.~\ref{fig_part_frac} (a) forms a regular 1D line (not a fractal line).

Following this, Partition-I is utilized to scrutinize the EE scaling for the model $H_{1}$ on $SC(n,1)$. As outlined in \textbf{Appendix}~\ref{app_scaling_b} and Fig.~\ref{fig_gap_gapless}, the energy gap of the model $H_{1}$ on the Sierpinski carpet diminishes at the scaling limit, and the DOS attains a finite value, indicating a metallic ground state on the fractal lattice. Concurrently, it is well-established that the model $H_{1}$ on a square lattice respects translation symmetry and displays a one-dimensional Fermi surface with a finite DOS. In accordance with the Widom conjecture of Toeplitz matrices, the EE scales as $S_{A} = a L_{A}\log L_{A} + \cdots$, where the coefficient $a$ is determined by the geometric details of the Fermi surface and the specifics of the partitioning method~\cite{gioevEntanglement2006}.

By comparing these two scenarios, we are motivated to test whether or not the EE when the model is placed on the Sierpinski carpet would also scale as $S_{A} = a L_{A}^{\alpha}\log L_{A} + \cdots$, where the parameter $\alpha$ might encode the fractal information of the lattice, and $a$ remains a nonuniversal constant determined by certain details of the model. To substantiate this proposal, we  numerically compute $S_{A}$ for different $n$ of $SC(n,1)$ to increase the fractal lattice size, where the largest size of fractal lattice $SC(5,1)$ that we calculated has $8^{5}$ lattice site, as depicted in Fig.~\ref{fig_part_frac} (b). By fitting the numerical data with $\alpha=1$, we obtain $S_{A}/ \log L_{A} = 0.119465 L_{A} + 0.609574$ with the \textit{coefficient of determination} $R^{2} = 0.999368$,  a statistical measure for the regression predictions approximating the real data points~\cite{devore2012probability}. Additionally, we utilize $R^{2}$ as a function of $\alpha$ to measure the goodness of fit, as shown in the inset of Fig.~\ref{fig_part_frac} (b). We observe that $\alpha\approx 1$ yields the best fit, with $R^{2}$ closest to $1$.  Based on this fitting expression in 2D where the fractal lattice is embedded, we propose that the EE scaling of the model $H_{1}$ on the fractal lattice is given by:
\begin{equation}
S_{A} = a L_{A}^{d_{s}-1} \log L_{A}+\cdots\,.
\label{eq_EE_scaling}
\end{equation}
Here, $d_{s}$ is the integer-valued spatial dimension of the Euclidean space where the fractal lattice is embedded. However, as only one specific partition scheme is applied during the above analysis, one may wonder if this scaling is universal enough.

To assess the universality of Eq.~\eqref{eq_EE_scaling} across different partition schemes, we incorporate Partition-II alongside Partition-I as depicted in Fig.~\ref{fig_part_frac}(a). In the case of Partition-II, $A$ explicitly breaks the symmetry and fractal structure of the original Sierpinski carpet. Usually, for regular lattice without fractal structure, the linear length $L_{A}$ of the boundary between subsystems $A$ and $B$ has $L_{A} = N_{bA}$.  Notably, here the linear length of $A$'s boundary is $L_{A}=3^{n-1} l$ in $SC(n,1)$ and the number of sites on the boundary of the subsystem $A$ is $N_{bA}=2^{n-1}$, illustrated in Fig.~\ref{fig_part_frac}(c) for varying $n$. Since $L_{A} > N_{bA}$, the boundary ``$FB(n,1)$'' of $A$ exhibits a distinct fractal structure, contrasting sharply with Partition-I where the boundary is a regular 1D line. To reflect the effect of the fractal structure of $A$'s boundary for the EE, without loss of generality, we choose $L_{A}=3^{n-1}$ as the linear length of the subsystem $A$. For convenience, we define a boundary Hausdorff dimension $d_{\rm bf}=\log_{L_{A}}N_{b{A}}=\log_{3}2$ for $FB(n,1)$, akin to the $n$th-order approximation of the Cantor set~\cite{mandelbrot1983fractal}.

With this foundation, we numerically determine the EE on $SC(n,1)$ as depicted in Fig.~\ref{fig_part_frac}(d). We anticipate that the EE scaling in this case remains $S_{A} = a L_{A}^{\alpha}\log L_{A} + \cdots$. Fitting the data with $\alpha=1$ yields $S_{A}/ \log L_{A}= 0.0837457 L_{A} + 1.15904$ with $R^{2}=0.998586$. The optimal fit is obtained with $\alpha\approx 1$, with the coefficient of determination $R^{2}$ closest to $1$, as shown in the inset of Fig.~\ref{fig_part_frac}(d). Based on these findings, the EE scaling of the model $H_{1}$ on the fractal lattice is consistent with Eq.~\eqref{eq_EE_scaling}, and breaking self-similar and symmetries of original fractal lattice do not affect the scaling behavior of EE. 
While it is impractical to exhaustively consider an infinite array of partition schemes, the two partitions discussed, i.e., Partition-I and Partition-II, which respectively preserve and break the symmetry of the original Sierpinski carpet, serve as representative cases. In \textbf{Appendix}~\ref{app_morescaling_c}, by adpoting Partition-III and Partition-IV, the scaling behavior of EE still satisfies Eq.~\eqref{eq_EE_scaling}. Thus, we conclude that Eq.~\eqref{eq_EE_scaling} universally governs the scaling of EE for gapless free-fermion systems with a finite DOS on a fractal lattice embedded in $d_s$ dimensional Euclidean space.  

Finally, noting that the EE of free-fermion systems can be formally expressed as $S_{A} = \text{Tr}[f(C^{A})]$ as shown in Sec.~\ref{section_defini}~\cite{peschelCalculation2003,klichLower2006}, Eq.~\eqref{eq_EE_scaling} provides information about the asymptotic behavior of the spectrum of the correlation matrix $C^{A}$. This may assist in analytically generalizing the Widom conjecture of Toeplitz matrices to matrices with self-similarity. Mathematically, Brownian motion on infinitely ramified self-similar fractals (see Refs.~\cite{rammalRandom1983,goldstein1987percolation} and references therein), such as the Sierpinski carpet, presents a challenging problem concerning the asymptotic behaviors of the Laplacian on the fractals~\cite{kajinoSpectral2010,bassUniqueness2010}. Formulating a new `conjecture' for the asymptotic behaviors of matrices with self-similarity would be instrumental in understanding Brownian motion on fractals, representing an important avenue for future study.

\begin{figure*}[htbp]
\centering
\includegraphics[width=17.8cm]{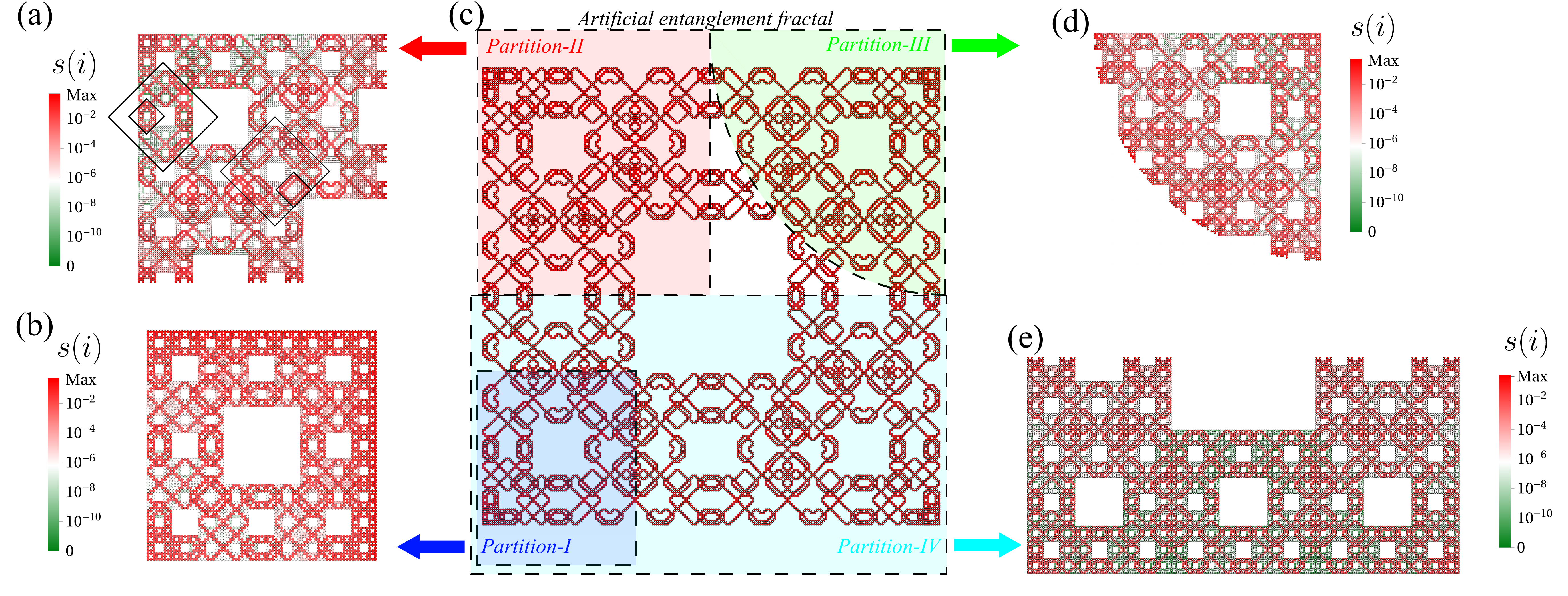}
\caption{(a), (b), (d) and (e) are EC of the model $H_{1}$ using Partition-II,I,III and IV, respectively.  (c) The artificially generated structure, called \emph{entanglement fractal} (EF), and its four partitions. The predominant  EC  data (formed by significantly large value of $s(i)$ whose color tends to be red) of (a), (b), (d), and (e) match the artificially generated EF pattern in (c). 
}\label{fig_entang_selfsimi}
\end{figure*}

\begin{figure}
\centering
\includegraphics[width=7.8cm]{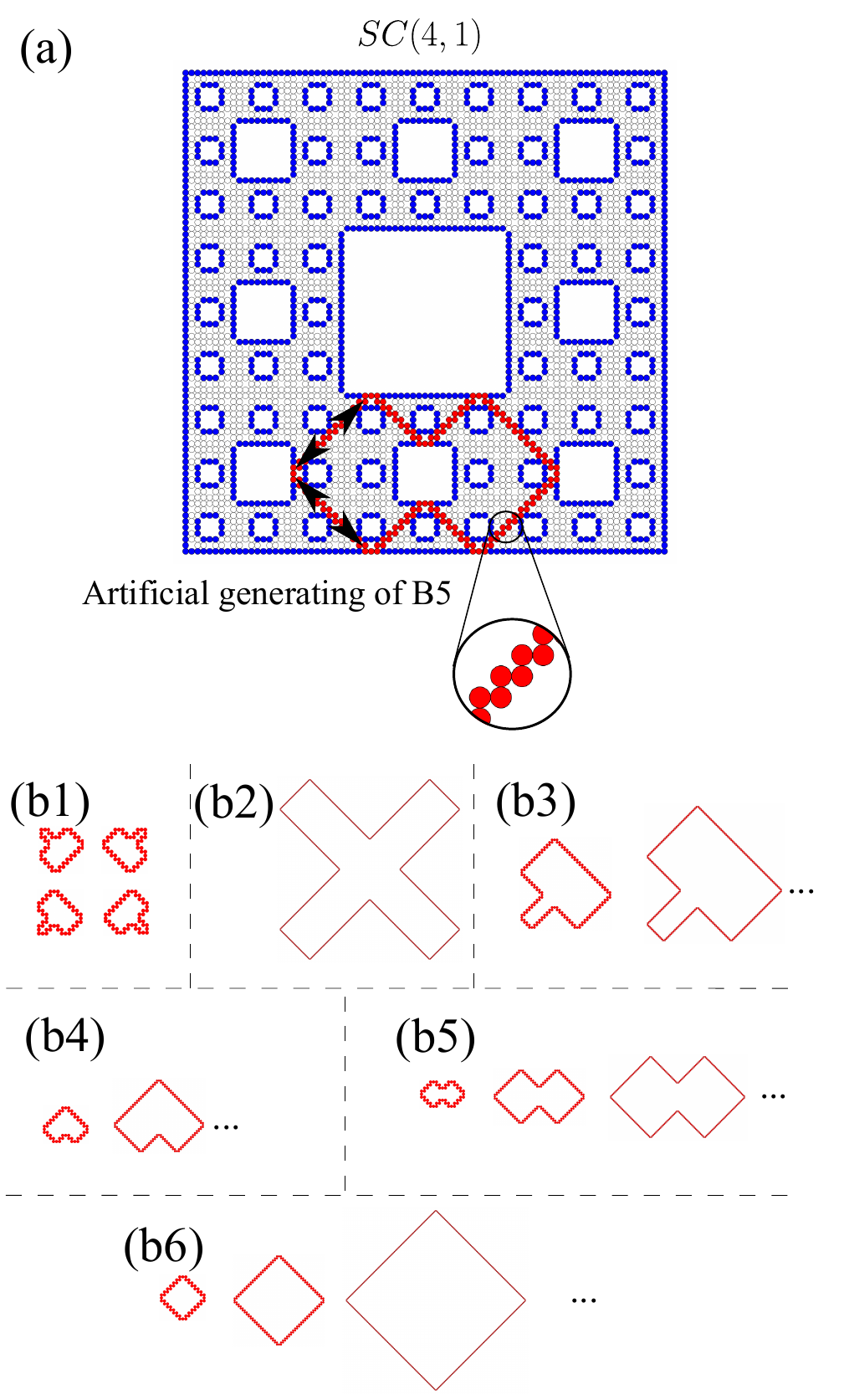}
\caption{(A) Illustration of the artificial generating rules for \emph{entanglement fractal} on the $4$th-order approximation $SC(4,1)$. The red closed loop shows the artificial generating structure of (B5) in the gapless fractal systems and the blue sites are the boundaries of $SC(4,1)$. (B1-B6) All possible structures by the artificial generating rules.}\label{fig_art_rule}
\end{figure}

\section{Entanglement fractal and its artificial generating}
To deeply probe into the entanglement characteristics of fractal geometry and the intricate details of EE, we embark on an investigation of a measure known as the entanglement contour (EC)~\cite{chenEntanglement2014}, denoted as $s(i)$ where $i$ represents any lattice site within the subsystem $A$. EC provides a real-space ``tomography'' of EE and is defined in Eq.~\eqref{eq_ec_def} (for more details, refer to Sec.~\ref{section_defini}). By utilizing EC, we can reconstruct EE as:
 \begin{align}
 S_{A} = \sum_{i\in A} s(i)\,.\label{eq_ee_decompose}
 \end{align} Initially, we employ Partition-II to examine the distribution of EC $s(i)$ for the model $H_{1}$ on $SC(5,1)$. As illustrated in Fig.~\ref{fig_entang_selfsimi}(a), our numerical findings indicate that the predominant  $s(i)$ in the scaling limit, whose color tends to be red, exhibits a distinct pattern within the bulk of the subsystem $A$,  reminiscent of Chinese papercutting. Additionally, we observe self-similarity in the EC, as indicated by the black rhombuses in Fig.~\ref{fig_entang_selfsimi}(a).
 To advance our exploration, we employ Partition-I, III, and IV to examine the EC of the model $H_{1}$ on $SC(5,1)$. All numerical results are presented in Fig.~\ref{fig_entang_selfsimi}(a), (b), (d), and (e), where the predominant  EC data (formed by significantly large value of $s(i)$ whose color tends to be red) display a clearly self-similar structure.

Up to this point, we haven't yet discussed the pattern in Fig.~\ref{fig_entang_selfsimi}(c). This pattern is referred to as ``entanglement fractal'' (EF), which is not obtained through numerical computations but is instead artificially generated using a set of rules. We will discuss EF below and discover that EF effectively represents the real-space distribution of the predominant  EC data observed in Fig.~\ref{fig_entang_selfsimi}(a), (b), (d), and (e) (as indicated by the four arrows from (c) to (a), (b), (d), and (e)).

More concretely, to comprehend the numerical outcomes, we delineate a set of generative rules to artificially construct predominant  EC on the fractal lattice $SC(n,1)$ illustrated in Fig.~\ref{fig_entang_selfsimi}(a), (b), (d), and (e). As an illustrative instance, we depict a red closed loop within $SC(4,1)$ in Fig.~\ref{fig_art_rule}(a), where the blue sites form the boundaries of $SC(4,1)$. The fundamental rules encompass:

\begin{enumerate}
\item[\textit{i.}] The predominant  EC is constituted by closed loops;
\item[\textit{ii.}] Each loop line consists of two sites arranged in a zig-zag pattern, as depicted in the magnified view in Fig.~\ref{fig_art_rule}(a).
\item[\textit{iii.}] Each loop line extends along one of the four diagonal directions: $45^{\circ}$, $135^{\circ}$, $225^{\circ}$, or $315^{\circ}$, as indicated by the black arrows in Fig.~\ref{fig_art_rule}(a);
\item[\textit{iv.}] Each loop line originates from and reflects specularly at the boundaries of $SC(n,1)$.
\end{enumerate} 
By adhering to these rules, we artificially generate all conceivable structures on the fractal lattice $SC(n,1)$, as shown in Fig.~\ref{fig_art_rule}(b1-b6). Notably, the patterns in Fig.~\ref{fig_art_rule}(b3-b6) exhibit self-similarity, while the pattern in Fig.~\ref{fig_art_rule}(b4) does not appear in the EC of $H_1$ on $SC(n,1)$. The pattern of EC may be influenced by the underlying fractal structure. Specifically, when only the unit cell is modified while maintaining the same fractal iteration method for generating the lattice, we observe that the EC with a finite density of states such fractal lattice retains a similar structure. In contrast, altering the fractal iteration method leads to changes in the EC with finite DOS on this fractal lattice (see \textbf{Appendix}~\ref{app_moreec_e}, and Fig.~\ref{fig_entang_con_sc} and Fig.~\ref{fig_entang_con_gsc} for more details). Additionally, these rules should have an intrinsic connection with the fractal iteration method, where our results may provide a clue to the potential mathematical proof of the connection.

By applying the patterns in Fig.~\ref{fig_art_rule}(b1-b3) and (b5-b6) to $SC(5,1)$, we successfully generate a fractal structure on $SC(5,1)$, as depicted in Fig.~\ref{fig_entang_selfsimi}(c). Subsequently, by employing Partition-(I-IV) to Fig.~\ref{fig_entang_selfsimi}(c) and contrasting the resulting patterns with the numerical outcomes in Fig.~\ref{fig_entang_selfsimi}(a), (b), (d), and (e), we observe that the distribution of predominant  EC data in the bulk of $A$ closely resembles the artificially generated pattern shown in Fig.~\ref{fig_entang_selfsimi}(c). This similarity indicates that the chosen rules have been successful. The distinctive fractal pattern, artificially generated by a set of rules, is referred to as the EF, which can be considered as the entanglement ``fingerprint'' of the fractal geometry of the original lattice.  Although the EF pattern may appear aesthetically pleasing, its physical significance and relevance to physics may be questioned. Compared with the EC of the gapped system in the Sec.~\ref{Sec_gap}, EF only appears in gapless systems with finite DOS, where we consider that EF plays a vital role in generalizing the super-area law from translationally invariant lattice to fractal lattice.

\section{Two distinct scaling behaviors in EC}
As shown in Eq.~\eqref{eq_ee_decompose}, the EE can be decomposed into the real-space distribution within the subsystem $A$. Meanwhile, we observed the entanglement fractal distinct from the fractal structure of Sierpinski carpet, which suggest that $s_{i}$ respectively on entanglement fractal and the rest of the sites in subsystem $A$ should have different scaling behaviors. Then, we will elaborate on this point in more detail below.

\begin{figure}[htbp]
\centering
\includegraphics[width=8.5cm]{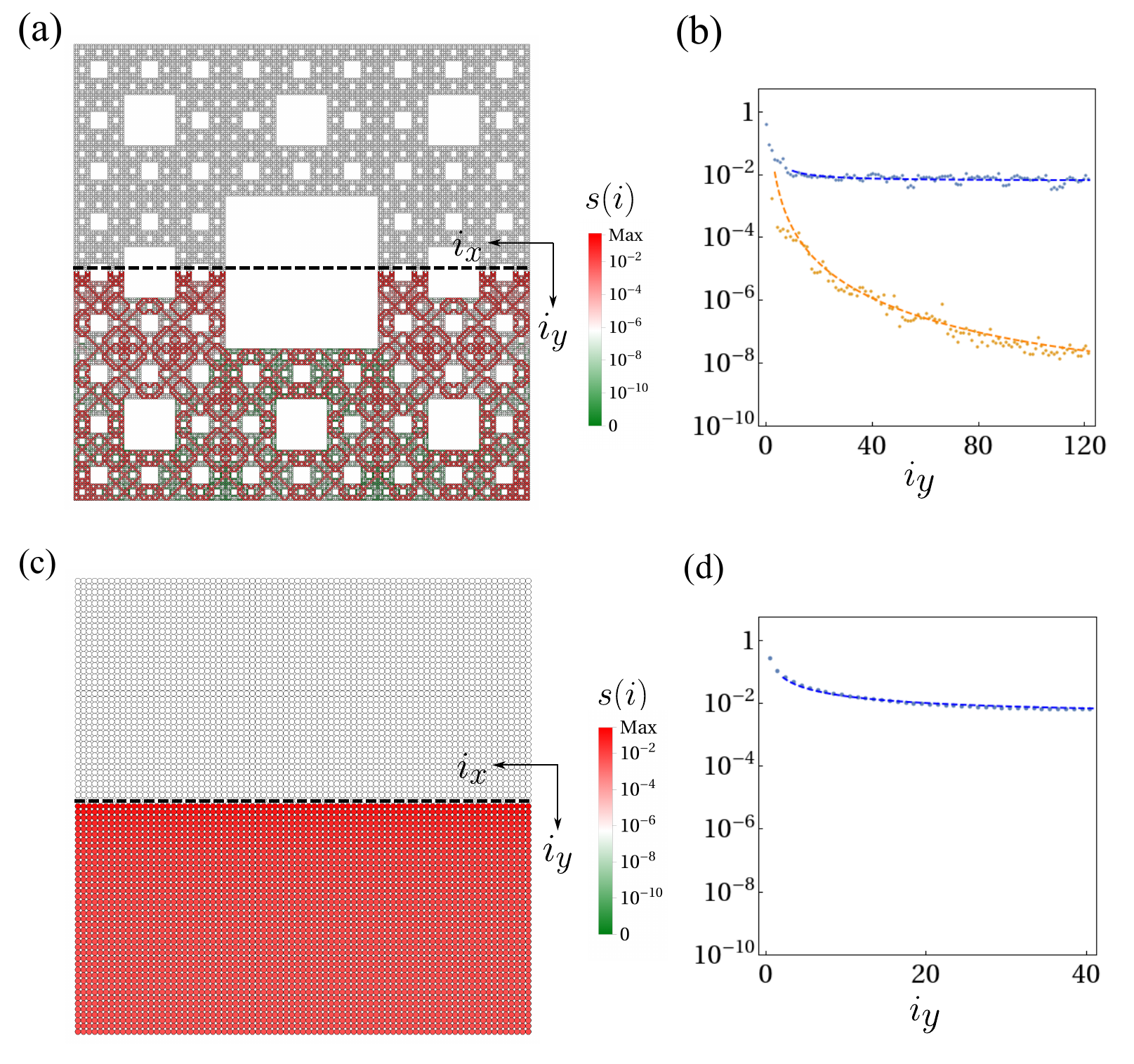}
\caption{(a) is the EC of the model $H_{1}$ using Partition-IV. (b) The behavior of EC $s(i)$ on the self-similar structure $A_{s}$ and the rest of the subsystem $\bar{A}_{s}$ by adpoting Partition-IV in (a), where the blue and yellow lines represent $s_{A_{s}}(i_{y})\sim 1/(i_{y})^{1.33}$ and $s_{\bar{A}_{s}}(i_{y})\sim 1/(i_{y})^{3.78}$, respectively. (c) is the EC of the model $H_{1}$ on square lattice by using Partition-IV. (d) EC on square lattice only has a single power-law scaling $s(i_{x},i_{y})\sim 1/i_{y}$.}\label{fig_EF_data}
\end{figure}

To explore the distinct behaviors of $s(i)$ on the EF and the remaining part of subsystem $A$, as illustrated in Fig.~\ref{fig_entang_selfsimi}(e) where Partition-IV is applied, we introduce two quantities defined as:
\begin{align}
s_{A_{s}(\bar{A}_{s})}(i_{y})= \frac{\sum_{i_{x}\in A_{s}(\bar{A}_{s})}s(i_{x},i_{y})}{N_{i_{y}}}\,,\label{eq_two_quant}
\end{align}
where $A_{s}$ and $\bar{A}_{s}$ represent the lattice sites covered by the EF and the rest of the sites in subsystem $A$, respectively. Each lattice site $i$ is uniquely labeled by two integers: $(i_x,i_y)$ in Fig.~\ref{fig_EF_data}(a). Here, $N_{i_{y}}$ is the number of lattice sites on the EF or the rest of sites in subsystem $A$ with the same $i_y$ in Fig.~\ref{fig_EF_data}(a). The coordinate $i_y$ measures the distance of lattice site $i$ from the boundary between the two subsystems $A$ and $B$.

From Eq.~\eqref{eq_two_quant},  $s_{A_{s}(\bar{A}_{s})}(i_{y})$  represents the average value of EC data for a given  $i_y$ across all sites  $i_x \in A_{s}(\bar{A}_{s})$. As depicted in Fig.~\ref{fig_EF_data}(b), we numerically find that as $ i_{y}$  increases:
\begin{align}
s_{A_{s}}(i_{y}) \sim \frac{1}{(i_{y})^{1.33}}, \quad s_{\bar{A}_{s}}(i_{y}) \sim \frac{1}{(i_{y})^{3.78}}.
\label{eq_numerical_twoqq}
\end{align}
The coexistence of these two power-law scaling in EC has never been observed in translation-invariant systems~\cite{chenEntanglement2014}. This coexistence indicates that when increasing the size of the fractal system, only the physical degrees of freedom on EF have the dominant contribution to physical quantities, such as EE. Consequently, EF can be regarded as an emergent structure in the renormalization process. Moreover, combing the EF and assuming the scaling of EC on EF  $s_{A_{s}}(i_{y})\sim \frac{1}{i_{y}}$, we can reconstruct the super-area law of EE as shown in Eq.~\eqref{eq_EE_scaling} (more details see \textbf{Appendix}~\ref{app_recons_d}). As shown in Eq.~\eqref{eq_numerical_twoqq}, the exponent of $s_{A_{s}}(i_{y})$ is not equal to $1$, where we believe that the deviation of the exponent is originated from the finite-size effect and the result in Fig.~\ref{fig_EF_data}(b) is the best result that we can obtain numerically. 

To compare with the EC of translation-invariant systems, we numerically calculate the EC of the model  $H_{1}$ on a square lattice using Partition-IV, as shown in Fig.~\ref{fig_EF_data}(c). We find that  $s(i_{x},i_{y})$  only depends on $ i_{y}$. We then plot the scaling behavior of EC with increasing  $i_{y}$ and fixed  $i_{x}$, as shown in Fig.~\ref{fig_EF_data}(d). We observe that the EC of the square lattice exhibits a single power-law scaling $s(i_{x},i_{y})\sim 1/i_{y}$.
Returning to the fractal system, and combing our numerical results for EE in Fig.~\ref{fig_part_frac}, we propose that the power-law scaling of EC is a necessary condition for the existence of super-area law in translation-invariant or fractal system. However, the coexistence of two distinct scaling behaviors in fractal systems need further theoretical study.

\begin{figure}[htbp]
\centering
\includegraphics[width=8.7cm]{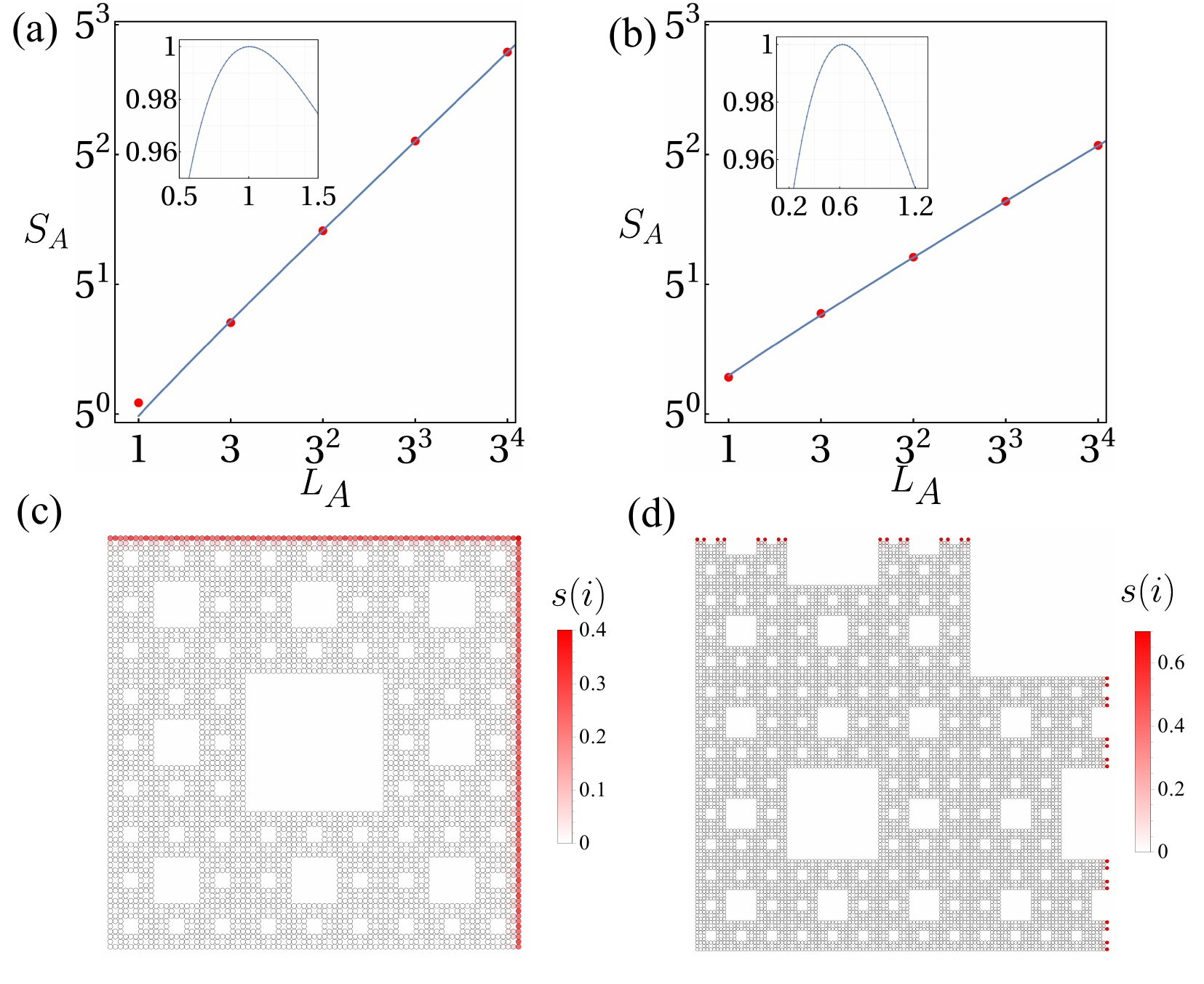}
\caption{(a) and (b) are the EE of the model $H_{2}$ as a function of the linear length $L_{A}$ of subsystem $A$ by using Partition-I and II, respectively. The insets show the coefficient of determination $R^{2}$ as a function of $\alpha$ to measure the goodness of fit. (c) and (d) are the distribution of EC of the model $H_{2}$ in the subsystem $A$ of $SC(5,1)$ by using Partition-I and  II, respectively. Here $t_{1}=0.5$ and $t=1$.}\label{fig_entang_gap}
\end{figure}

\section{EE \& EC of gapped ground states}\label{Sec_gap}

Turning to the entanglement of gapped systems on fractals,  we consider a tight-binding model on the fractal lattice $SC(n,1)$ embedded in a two-dimensional space, as shown in Fig.~\ref{fig_part_frac}(a). The model is given by
\begin{equation}
H_{2}=\sum_{i}(c^{\dagger}_{s,i} c^{\dagger}_{p,i}) t_{1} \sigma_{x}
\begin{pmatrix}
    c_{s,i} \\
    c_{p,i} 
\end{pmatrix}
+ \sum_{<i,j>}  ( c^{\dagger}_{s,i}  c^{\dagger}_{p,i})t\sigma_{z}
\begin{pmatrix}
    c_{s,j} \\ 
    c_{p,j}
\end{pmatrix},
\end{equation}
where $c^{\dagger}_{s(p),i(j)}$ is a fermionic creation operator of $s(p)$ orbital at the $i(j)$th lattice site and $\sigma_{x,y}$ are Pauli matrices. As discussed in \textbf{Appendix}~\ref{app_scaling_b} and Fig.~\ref{fig_gap_gapless}, through analyzing the scaling of the energy gap and the DOS, we find that the ground state of the model is gapped.

Due to the existence of a finite gap, we propose that the EE of this model $H_{2}$ on the Sierpinski carpet would scale as $S_{A}=aL_{A}^{\alpha}+\cdots$, where $\alpha$ is a universal parameter to be determined. To verify this proposal,  we first adopt Partition-I in Fig.~\ref{fig_part_frac}(a) to study the scaling of EE of the model $H_{2}$ on the $n$th-order approximation $SC(n,1)$ of Sierpinski carpet. As shown in Fig.~\ref{fig_entang_gap}(a), by fitting the data with $\alpha=1$, the numerical data of EE is fit to $S_{A} = 1.10033L_{A}-0.120981$. The best fit is $\alpha\approx 1$ with the coefficient of determination $R^{2}$ closest to $1$, as shown in the inset of Fig.~\ref{fig_entang_gap}(a). 
Furthermore, we consider Partition-II in Fig.~\ref{fig_part_frac}(a) to divide $SC(n,1)$ into $A$ and $B$, where the boundary $FB(n,1)$ in the blue area of Fig.~\ref{fig_part_frac}(a) has fractal structure. Through numerical calculation, the EE of the model $H_{2}$ in this case is demonstrated in Fig.~\ref{fig_entang_gap}(b). By fitting the numerical data, the EE scales to $S_{A}=1.84893L_{A}^{0.62}-0.234274$, where $\alpha\approx 0.62$ is the best fit with the coefficient of determination $R^{2}$ closest to $1$. Note that for Sierpinski carpet, its spatial dimension $d_{s}=2$ and Hausdorff dimension $d_{f}=\log_{3}8\approx 1.8928$, while the boundary $FB(n-1)$ of the subsystem $A$ has the Hausdorff dimension $d_{\rm bf}=\log_{3}2\approx 0.6309$. Combining the numerical results of EE with Partition-I and II, we observe that the EE of the gapped systems  can reflect the fractal feature of the subsystem's boundaries. Then, for gapped systems on fractal, we obtain a generalized area law written as 
\begin{equation}
S_{A} = a L_{A}^{d_{\rm bf}} + \cdots,
\label{eq_EE_gap}
\end{equation}
where $d_{\rm bf}$ is the boundary Hausdorff dimension. Finally, considering the EC $s(i)$ in this case, as demonstrated in Fig.~\ref{fig_entang_gap}(c) and (d), the distribution of $s(i)$ is localized with uniformly exponential decay at the boundaries of the subsystem $A$.

\section{Discussions and outlook}
In this study, we have investigated the interaction between fractal geometry and quantum entanglement of free fermions using two entanglement measures: entanglement entropy (EE) and entanglement contour (EC). Several intriguing questions arise, prompting further exploration. For instance, we have numerically confirmed the existence of super-area law in the gapless system with finite DOS on fractal lattice. Furthermore, we have introduced several guidelines for artificially generating entanglement fractal (EF) as depicted in Fig.~\ref{fig_entang_selfsimi}(c), which emerge from the predominant EC data at the scaling limit. Finally, we observed the coexistence of two distinct scaling behaviors in EC of fractal systems, which have never be observed in translation-invariant systems.  

However, the foundational theory of EF remains to be established, representing a critical next step in comprehending the two assumptions outlined in the main text. To formulate this theory, one approach could involve continuously adjusting the hopping energies of a tight-binding model. This adjustment would lead to a gradual evolution of the hopping energy distribution from a regular translationally invariant lattice to a fractal lattice. By examining the changes in EE and EC throughout this evolution, we suggest that the topology of the Fermi surface~\cite{tamTopological2022} on the regular translationally invariant lattice could significantly influence the EF structure on the fractal lattice. Besides, we also wonder when we change the fractal structure of the lattice, whether EC still has the coexistence of two scaling behaviors. 

Furthermore, drawing inspiration from the Widom conjecture and analytical findings~\cite{leeFreefermion2015,leePositionmomentum2014} concerning Toeplitz matrices in translation-invariant systems for computing EE, it becomes imperative to investigate the asymptotic behavior of the correlation matrix $C^A$ on fractals. Such exploration could potentially aid in determining the analytical expression of the non-universal coefficient in the EE scaling formulas. Additionally, similar analyses can be conducted on hyperbolic lattices, where insights from non-Abelian hyperbolic band theory may prove beneficial~\cite{maciejkoAutomorphic2022,HuangEntanglement2024}. Furthermore, for exactly solvable models of interacting many-body systems exhibiting exotic scaling of  EE~\cite{movassaghSupercritical2016}, exploring the scaling of their EC presents an intriguing avenue, potentially revealing universal behaviors extending beyond free-fermion systems. In addition, with the discovery of exotic entanglement phenomena in non-Hermitian systems, the definition and the universal feature of EC in such systems is an interesting direction for further exploration~\cite{Chenquantum2024}. Fractal structure often appears at phase transition points, such as the phase transition point of $2$D Ising model~\cite{itoFractal1987}. it is a interesting question to study the effect of this emergent fractal structure at phase transition points for the property of quantum entanglement.

The correlation matrix typically retains symmetries inherited from the Hamiltonian of lattice systems, including properties like translation invariance and self-similarity. By focusing on conjectures regarding the asymptotic behavior of the correlation matrix on fractals, we stand to glean valuable insights into the asymptotic tendencies of a particular matrix class exhibiting self-similarity. Moreover, such endeavors pave the way for examining Brownian motion on fractals, establishing connections with the spectral characteristics of the Laplacian on fractals~\cite{kajinoSpectral2010,bassUniqueness2010}.

With advancements in experimental techniques, it has become feasible to experimentally realize lattice systems with fractal structures in physics and chemistry~\cite{shangAssembling2015a,liConstruction2017,kempkesDesign2019a,moOnSurface2019,liuSierpi2021}. This development opens avenues for studying many-body systems with fractal geometries. Furthermore, in the realm of phononic platforms, it is now possible to simulate and measure the entanglement of many-body systems using pumping-probe responses in fractal phononic lattices~\cite{liHigherorder2022a,zhengObservation2022a,linMeasuring2024}. For free-fermion systems, the entanglement contour can be decomposed into the summation of particle-number cumulants' densities, providing a method to measure the entanglement fractal in the transport of quantum point contacts~\cite{moUnraveling2023}.

\emph{Acknowledgments}---
P.Y. and Y.Z. were supported by the National Natural Science Foundation of China under the grant No. 12074438, the Guangdong Basic and Applied Basic Research Foundation under the grant No. 2022B1212010008


%

\bigskip

\bigskip

\appendix
\begin{center}
{  \textbf{Appendix}}
\end{center}

\section{Fractal lattice and the definition of entanglement quantity}\label{app_fractal_a}
\begin{figure}
\centering
\includegraphics[width=8.5cm]{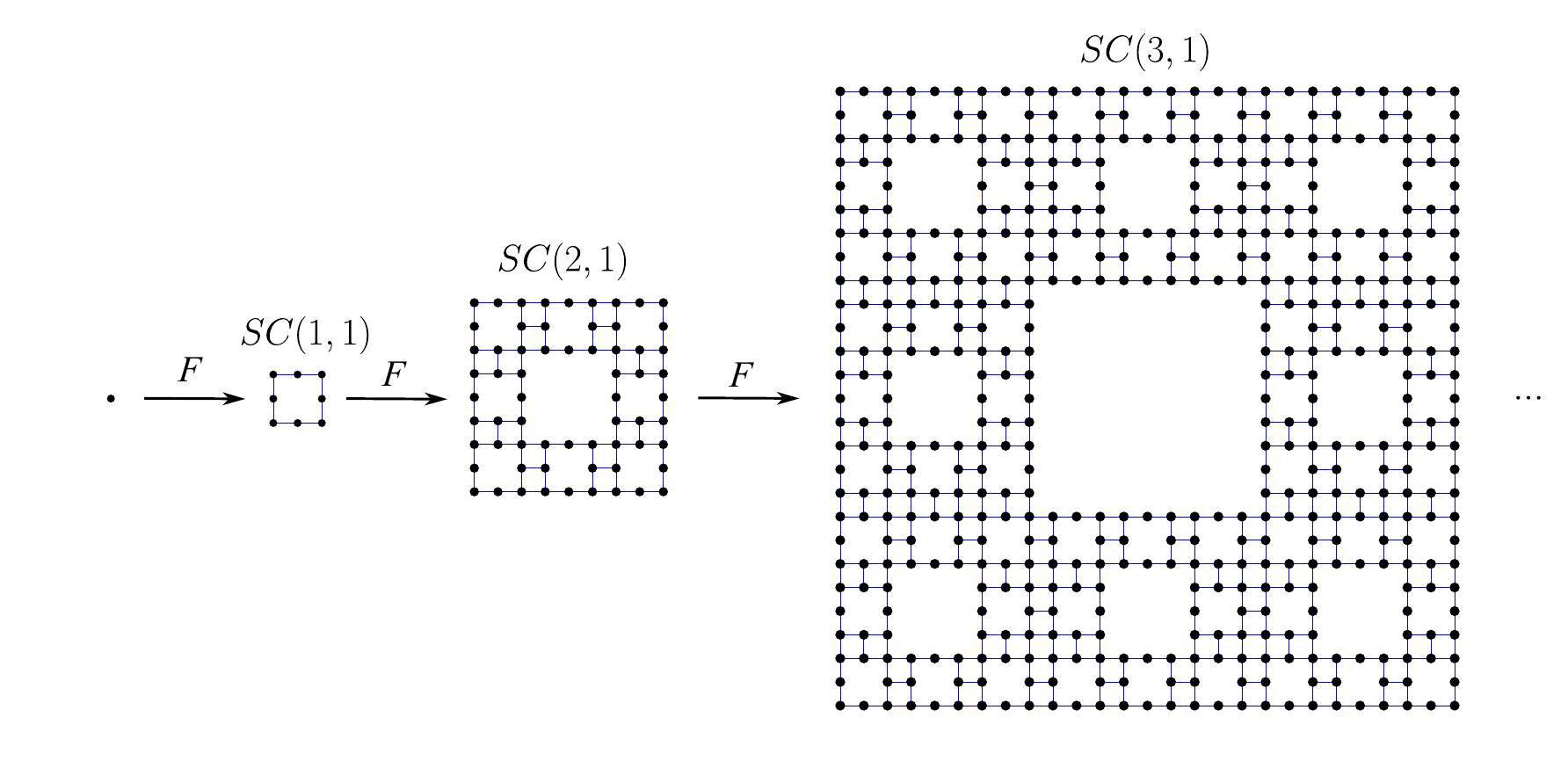}
\caption{Illustration of using iteration method $F$ to generate the $n$th-order approximation $SC(n,1)$ of Sierpinski carpet with $N=8^{n}$ lattice sites.}\label{fig_frac_iter}
\end{figure}

Here, we discuss the construction of a lattice system with a fractal structure. Initially, we consider an initial unit cell $U$ and a fractal iteration method $F$ to generate the fractal lattice. As illustrated in Fig.~\ref{fig_frac_iter}, employing the method $F$ on the cell $U$ iteratively $n$ times allows us to obtain the $n$th-order approximation $SC(n, s)$ of the Sierpinski carpet~\cite{mandelbrot1983fractal}. Here, $n$ represents the $n$th iteration, and the number of lattice sites in a unit cell is $s^{2}$. Without loss of generality, when $n \rightarrow \infty$ and setting the lattice constant $l=1$, the Hausdorff fractal dimension of the Sierpinski carpet is defined as
\begin{equation}
d_{f} = \lim_{n\rightarrow \infty}\log_{\mathcal{L}}\mathcal{N} = \log_{3}8,
\end{equation}
where the number of lattice sites $\mathcal{N} = s^{2} \times 8^{n} $ in the $n$th-order approximation $SC(n,s)$ of the Sierpinski carpet, and the width of $SC(n,s)$ is $\mathcal{L}= s \times 3^{n}$. The Hausdorff fractal dimension $d_{f}$ does not depend on the number $s^{2}$ of lattice sites in a unit cell $U$ when $n\rightarrow\infty$. Then, for convenience, we set $s=1$ in this work.

\begin{figure}
\centering
\includegraphics[width=8.5cm]{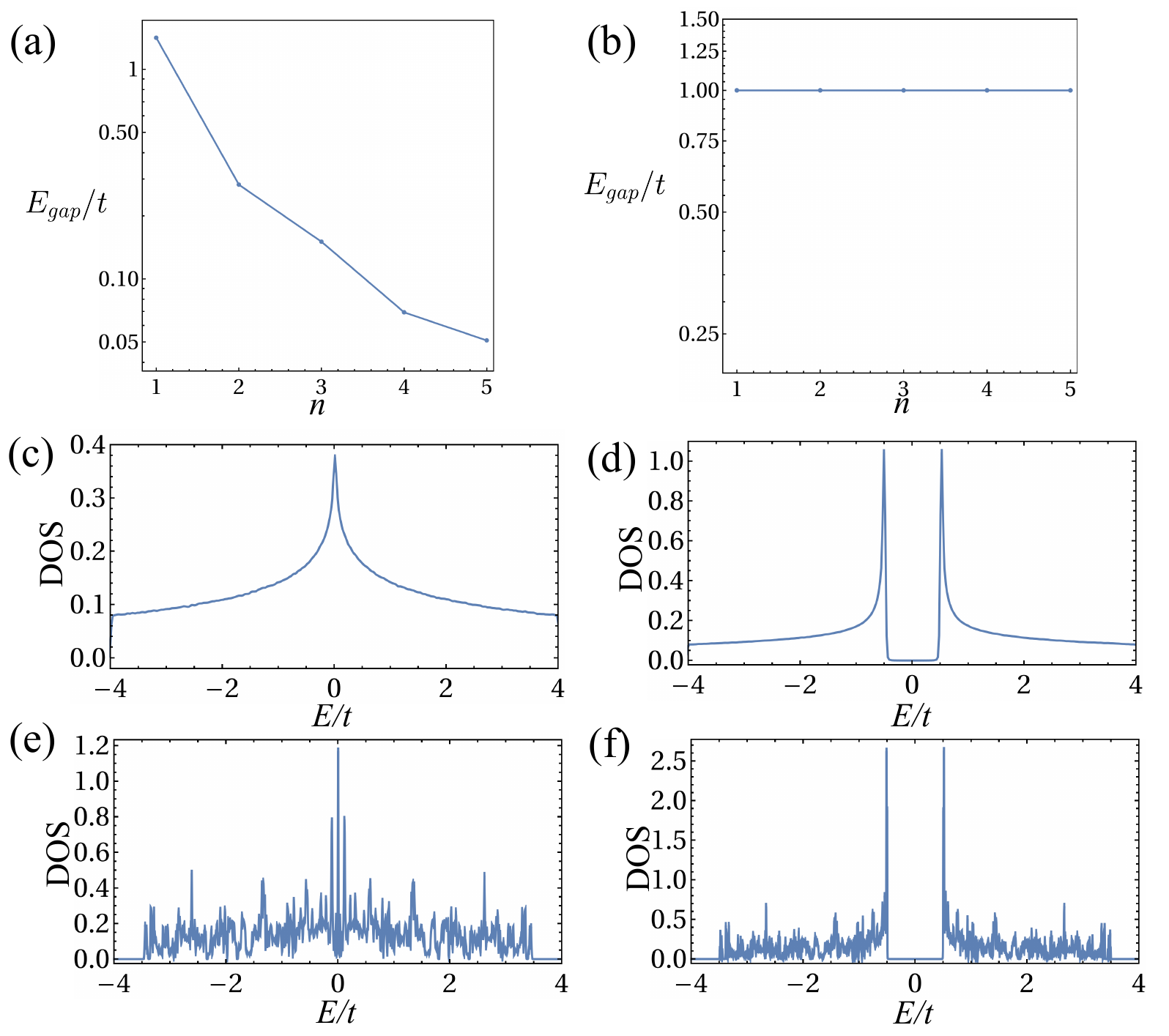}
\caption{Scaling of maximum energy gap $\text{max}(E_{j})$ for the model and  in (a) and (b).  
(c) and (d) are the DOS of the model and  on a square lattice with periodic boundary condition, respectively.  (e) and (f) are the DOS of the model and  on the approximation $SC(6,1)$ of Sierpinski carpet, respectively. Here $t=1$ and $\mu=0$ in the model, $t=1$, $t_{1}=0.5$ in the model. }\label{fig_gap_gapless}
\end{figure}

\section{Scaling of energy gap and DOS of the model \(H_{1}\) and \(H_{2}\) on fractal \(SC(n,1)\)}\label{app_scaling_b}
In this section, we turn into the properties of the energy spectrum for the models $H_{1}$ and $H_{2}$. For the $n$th-order approximation $SC(n,1)$ with a finite number of lattice sites, its energy spectrum exhibits a finite number of gaps ${E_{j}}$. We then examine the scaling behavior of the maximum energy gap $\text{max}(E_{j})$ as the number $n$ of iterations increases. As illustrated in Fig.~\ref{fig_gap_gapless}(a) and (b), we observe that the scaling of $\text{max}(E_{j})$ for model $H_{1}$ decreases exponentially, while $\text{max}(E_{j})$ for model $H_{2}$ remains invariant.

Furthermore, we consider the density of states (DOS) of the models $H_{1}$ and $H_{2}$ on $SC(n,1)$ to provide more information about their energy spectrum. Specifically, when model $H_{1}$ is defined on the square lattice with translation invariance, lattice constant $l=1$, and periodic boundary conditions, we can utilize Bloch band theory to clearly determine its energy spectrum. Moreover, its density of states in the thermodynamic limit can be obtained from the energy dispersion:
\begin{equation}
E(\bm{k}) = -2t(\cos k_{x}+ \cos k_{y}) - \mu,
\end{equation}
where the DOS in Fig.~\ref{fig_gap_gapless}(c) is depicted with $t=1$ and $\mu=0$. From the continuity of the DOS in Fig.~\ref{fig_gap_gapless}(c), we observe that the energy spectrum of the model on the square lattice is gapless, and the maximum point of the DOS indicates the presence of a nesting Fermi surface.

Next, we consider model $H_{1}$ defined on the $n$th-order approximation of the Sierpinski carpet. To practically determine the energy gap of the model on $SC(n,1)$, we adopt the method outlined in Ref.~\cite{yuanModeling2010} to calculate its DOS. This method proves to be very useful and efficient for large lattice systems without translation invariance. Utilizing this approach, we demonstrate the DOS of the model on the $6$th-order approximation $SC(6,1)$ in Fig.~\ref{fig_gap_gapless}(e) as an example, where $SC(6,1)$ comprises $8^{6}$ lattice sites. Upon comparison with the DOS in Fig.~\ref{fig_gap_gapless}(c), we observe that although the fractal structure induces fluctuations in the DOS of the model, it remains continuous, with its maximum point still located at $E=0$. Furthermore, with an increase in the number $n$ of iterations, the maximum energy gap $\text{max}(E_{j})$ of the model on $SC(n,1)$ becomes progressively smaller. Therefore, we propose that the model on the Sierpinski carpet, in the thermodynamic limit, can be regarded as a gapless system.

Here, we consider the DOS of model $H_{2}$ to determine the nature of its energy spectrum. Firstly, when considering model $H_{2}$ on the square lattice with periodic boundary conditions, we can determine its DOS in the thermodynamic limit using Bloch band theory, which is determined by the energy dispersion:
\begin{equation}
E(\bm{k})=\pm \sqrt{4t^{2}(\cos k_{x}+\cos k_{y})^{2}+t_{1}^{2}}\,,
\end{equation}
where the DOS is illustrated in Fig.~\ref{fig_gap_gapless}(d) with $t=1$ and $t_{1}=0.5$. We observe that the energy spectrum of the model in this case exhibits an energy gap. Next, considering the model defined on the approximation $SC(6,1)$ of the Sierpinski carpet as an example, by employing the efficient method outlined in Ref.~\cite{yuanModeling2010}, we find that the energy gap of the model still persists, as shown in Fig.~\ref{fig_gap_gapless}(f). Additionally, compared with the results in Fig.~\ref{fig_gap_gapless}(d), the fractal structure of $SC(6,1)$ induces fluctuations in the DOS and increases the DOS at the maximum point, as depicted in Fig.~\ref{fig_gap_gapless}(f). Furthermore, by increasing the number $n$ of iterations, the maximum energy gap $\text{max}(E_{j})$ of model $H_{2}$ on $SC(n,1)$ remains invariant, as shown in Fig.~\ref{fig_gap_gapless}(b). Therefore, we propose that the model on the Sierpinski carpet in the thermodynamic limit is still a gapped system.

\section{The scaling of EE by adopting Partition-III and IV}\label{app_morescaling_c}
To demonstrate how the boundary of the partition affects the scaling of EE, we study the scaling of EE using Partition-III and Partition-IV. Fitting the data with $\alpha=1$, we obtain $S_{A} / \log L_{A} = 0.0942769 L_{A} + 1.18486$ with $R^{2}=0.998813$ for Partition-III, and $S_{A} / \log L_{A} = 0.108694 L_{A} + 0.972181$ with $R^{2}=0.996532$ for Partition-IV. Additionally, we find that the optimal fit is achieved with $\alpha \approx 1$ in both Partition-III and Partition-IV. 
These results indicate that the boundary fractal structure of the partitions does not affect the scaling of gapless systems with finite DOS on the Sierpinski carpet, which is in contrast to the behavior observed in gapped systems.

\begin{figure}[htbp]
\centering
\includegraphics[width=8.6cm]{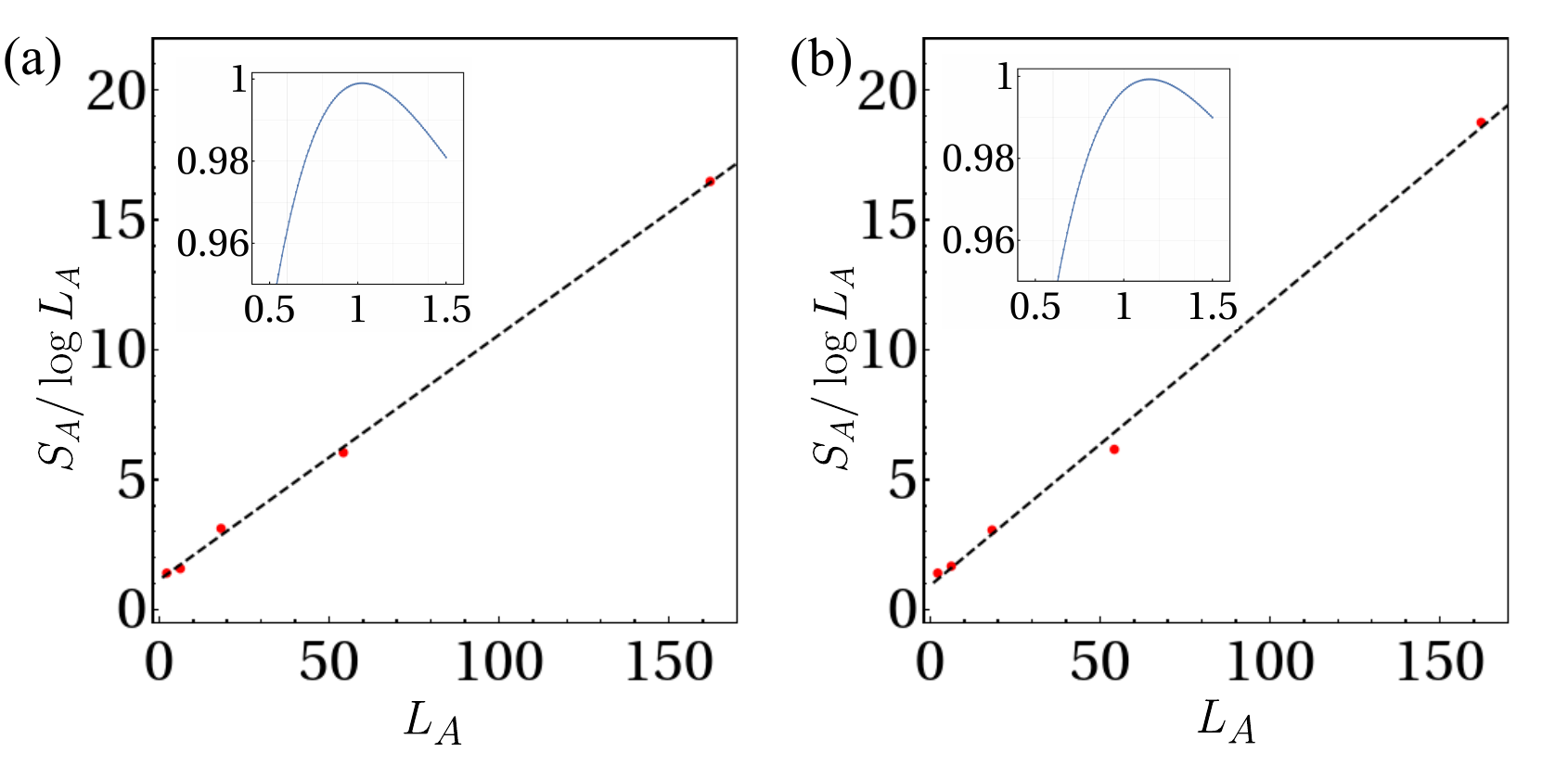}
\caption{The EE of the model on the $n$th-order approximation $SC(n,1)$ of the Sierpinski carpet using (a) Partition-III and (b) Partition-IV. The dashed line represents the linear fit of the EE data.}
\label{fig_part_ee_more}
\end{figure}

\section{More numerical results for entanglement contour}\label{app_moreec_e}
\begin{figure*}
\centering
\includegraphics[width=14.5cm]{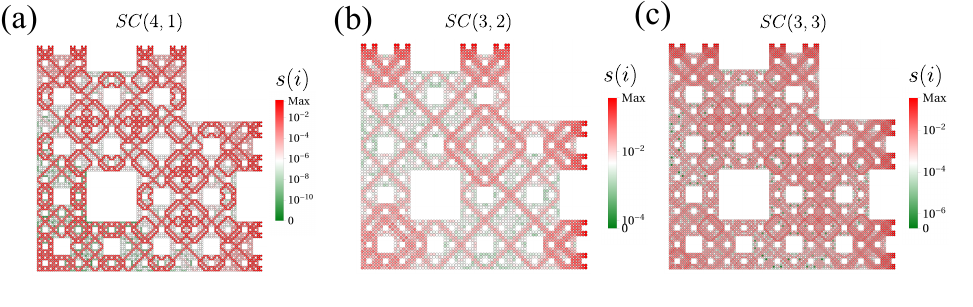}
\caption{(a-c) is the EC  of the model $H_{1}$ on the $n$th-order approximation $SC(n,s)$ with different unit cell. Here $t=1$ and $\mu=0$.}\label{fig_entang_con_sc}
\end{figure*}

For fractal systems, the fractal dimension is determined by the iteration method $F$ and the unit cell $U$. As the number of iterations $n$ approaches infinity, the fractal dimension solely depends on the iteration method $F$. Therefore, it is imperative to study the entanglement contour(EC) of model $H_{1}$ on a fractal lattice with identical iteration method but different unit cells to elucidate the entanglement fingerprint of fractal geometry. Specifically, in Fig.~\ref{fig_entang_con_sc}, we present numerical results of the EC of model $H_{1}$ in three types of approximations $SC(n,s)$ of the Sierpinski carpet, where the unit cell $U$ comprises $s^2$ lattice sites. Remarkably, we observe that for model $H_{1}$, its EC exhibits a universal pattern across the three types of approximations $SC(n,s)$, indicating that it is not influenced by the unit cell $U$. These numerical findings suggest that the structure of EC may solely depend on the iteration method.

\begin{figure*}
\centering
\includegraphics[width=14.5cm]{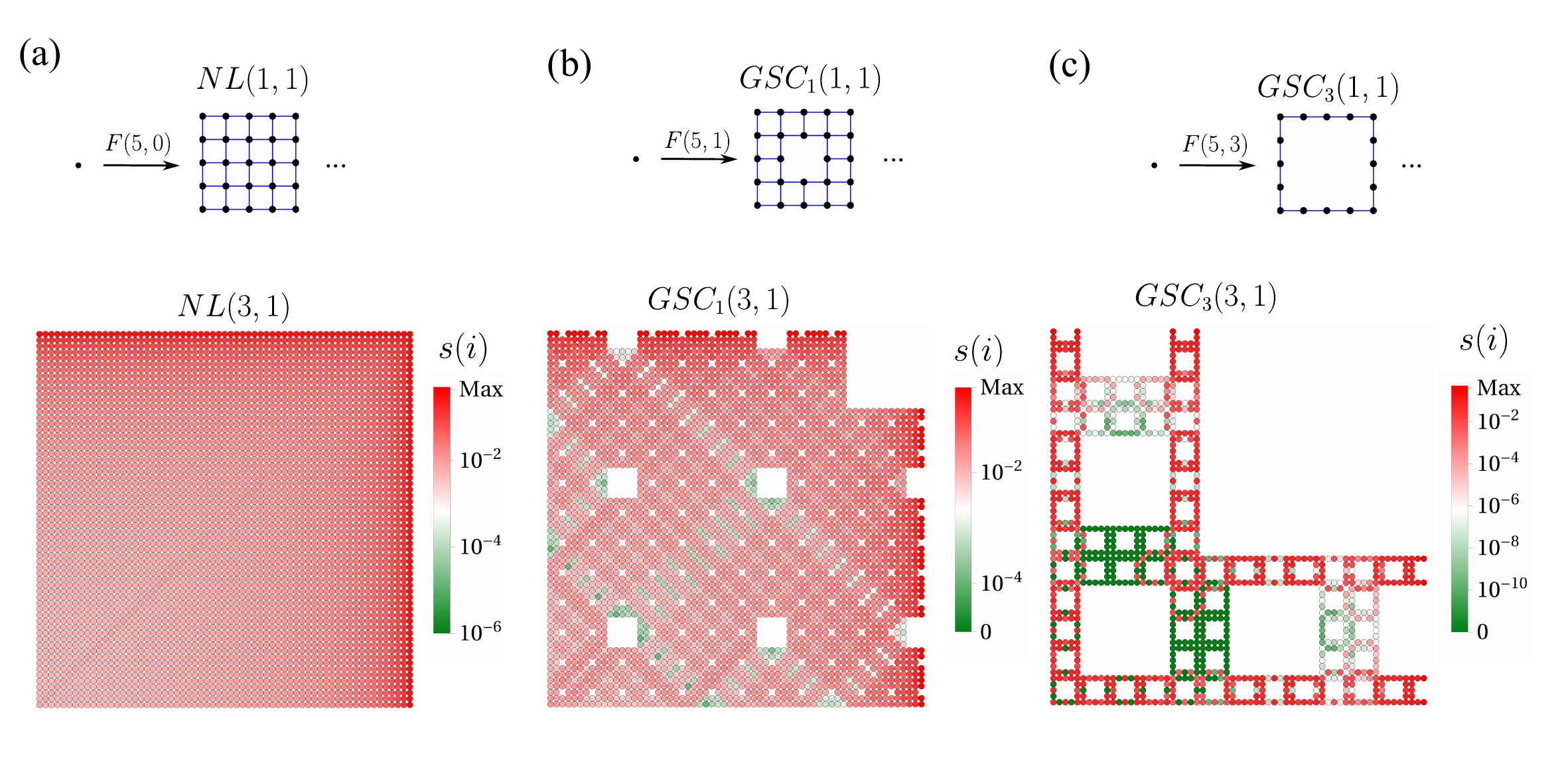}
\caption{ The upper of (a-c) show the generating process of normal lattice $NL(n,1)$ and the approximation $GCS_{1(3)}(n,1)$ generalized Sierpinski carpet. The bottom of (a-c) is the EC of the model $H_{1}$ on the normal lattice and the approximation $GSC_{1(3)}(3,1)$ of generalized Sierpinski carpet. Here $t=1$ and $\mu=0$.}\label{fig_entang_con_gsc}
\end{figure*}

In the following, we consider the effect of the iteration method $F$ on the EC of model $H_{1}$. Firstly, we provide a detailed discussion of the generalized Sierpinski carpet. As depicted in Fig.~\ref{fig_entang_con_gsc}, considering a unit cell with one lattice site, we employ an iteration method $F(m,m_{f})$ to act on the unit cell once. Then, the system comprises $m^2 - m_{f}^{2}$ lattice sites with a width of $m$ unit cells. Here $m > m_{f}$ and both are positive integer. If $m_{f}=0$, we obtain the normal lattice system $NL(n,1)$ with a trivial fractal structure and fractal dimension $d_{f}=2$, as shown in Fig.~\ref{fig_entang_con_gsc} (a), where $n$ is the number of iterations. When $m_{f}\neq0$, we obtain the $n$th-order approximation $GSC_{m_{f}}(n,1)$ of the generalized Sierpinski carpet in Fig.~\ref{fig_entang_con_gsc}(b) and (d), with the fractal dimension represented as:
\begin{equation}
d_{f}=\log_{m}(m^2-m_{f}^{2}). 
\end{equation}

Next, we investigate the EC of model $H_{1}$ on the approximation $GSC_{m_{f}}(n,1)$ of the generalized Sierpinski carpet. To eliminate irrelevant effects on the EC, we set $n=3$ to ensure that three lattice systems have identical widths, as shown in Fig.~\ref{fig_entang_con_gsc}. From the numerical results of the EC depicted in Fig.~\ref{fig_entang_con_gsc}(a-c), we observe that for the normal lattice $NL(n,1)$, the EC does not exhibit the special pattern consistent with the results in Ref.~\cite{chenEntanglement2014}. However, for the generalized Sierpinski carpet, the EC is influenced by the iteration method $F$, as shown in Fig.~\ref{fig_entang_con_gsc}(b-c). Based on these numerical findings, we conjecture that the EC of gapless systems with fractal geometry may exhibit a correspondence with the iteration method.

\section{Reconstruct the scaling of EE via entanglement fractal}\label{app_recons_d}
In this part, we use the entanglement fractal and the scaling of EC to reconstruct the scaling of EE.
Based on the definition of EC, we have this expression:
\begin{equation}
S_{A} = \sum_{i \in A_{s} }s(i) + \cdots.
\label{eq_fractEE}
\end{equation}
Here, the ellipsis represents the subleading term $\sum_{i\in \bar{A}_{s}} s(i)$ of the EE. Only the first term of  significantly contributes to the EE at the scaling limit. In terms of the quantity $s_{A_{s}}(i_{y})$, we have
\begin{align}
S_{A} \approx \sum_{i \in A_{s} }s(i)= \sum_{i_y} \mathcal{N}_{i_y} s_{A_{s}}(i_{y}).
\end{align} 
Since numerically we find that the denominator $\mathcal{N}_{i_{y}}$ in \eqref{eq_two_quant} is always less than $L_A$, we can rewrite the denominator as $\mathcal{N}_{i_{y}} = L_{A} \cdot p(i_{y})$ with $p(i_{y}) \equiv \mathcal{N}_{i_{y}}/L_{A}$ and $0<p(i_{y})<1$. Then we obtain:
\begin{equation}
S_{A} \approx L_A\sum_{i_y} p(i_{y}) s_{A_{s}}(i_{y})\,.
\label{eq_fractEE222}
\end{equation}
Here the series $\sum_{i_{y}}p(i_{y}) s_{A_{s}}(i_{y})$ can be regarded as a random series whose convergence depends on the distribution of $p(i_{y})$ and the series $\sum_{i_{y}}s_{A_{s}}(i_{y})$~\cite{durrett2019probability}. 

To reconstruct the super-area law of the EE, let us consider the linear size of $A$ is $L_A$ along both horizontal and vertical directions, i.e., $i_y\in (0, L_A)$, such that the subsystem is properly scaled to a thermodynamical two-dimensional region.
\begin{figure}[htbp]
\centering
\includegraphics[width=7cm]{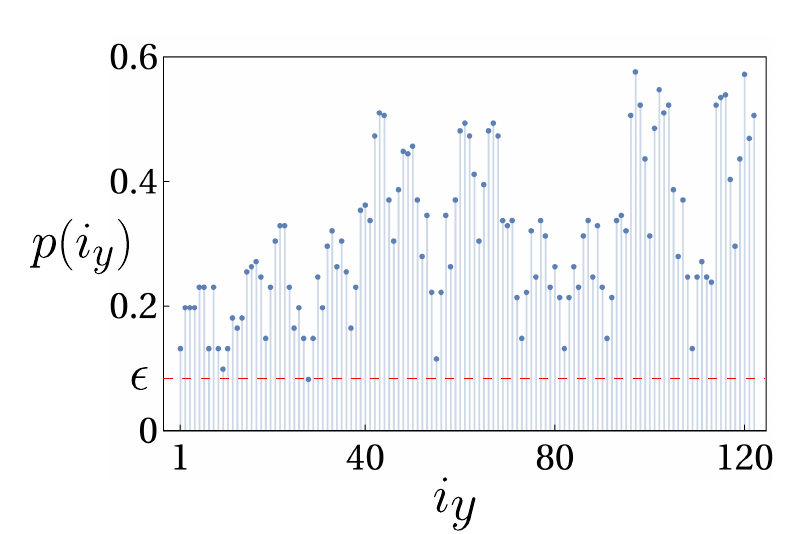}
\caption{ The distribution of $p(i_{y})$ with $i_{y}$ in the EC of Partition-IV, where $\epsilon$ is the lower bound of $p(i_{y})$. }\label{fig_EF_possi}
\end{figure}
Mathematically, the series $\sum_{n=1}^{\infty}1/n^{\beta}$ converges to a constant called the Riemann zeta function $\zeta(\beta)$ if $\beta > 1$. Consequently, when $s_{A_{s}}(i_{y})\sim 1/(i_{y})^{\beta}$ with $\beta>1$ is utilized, we can deduce the series $\sum_{i_{y}}p(i_{y}) s_{A_{s}}(i_{y})<\sum_{i_{y}}  s_{A_{s}}(i_{y})$ is convergent by noting that $0<p(i_{y})<1$. As a result, based on \eqref{eq_fractEE222}, we can deduce that the EE $S_{A} \lesssim  L_{A}$, which corresponds to, \textit{at most}, the area law but not the super-area law at large $L_A$ limit.

Therefore, to recover the logarithmic correction $\log L_A$ appeared in the super-area law in \eqref{eq_EE_scaling},  firstly,  we assume the following   asymptotic behavior:
\begin{align}
s_{A_{s}}(i_{y})\sim 1/i_{y}\label{eq_conjecture_s_as}\,.
\end{align}
Secondly,  we assume that $p(i_{y})$ should have a non-zero lower bound $p(i_{y})>\epsilon$. As shown in Fig.~\ref{fig_EF_possi}, we plot $p(i_{y})$ and numerically confirm that the nonzero lower bound of $p(i_{y})$ indeed exists. Based on these two assumptions, and combining \eqref{eq_fractEE222} and \eqref{eq_conjecture_s_as}, we can infer the EE should adhere to $\epsilon L_A\log L_{A} \lesssim S_{A} \lesssim L_{A} \log L_{A} $, i.e., $S_{A} \sim L_{A} \log L_{A} $ at large $L_A$ limit, which aligns with the super-area law in \eqref{eq_EE_scaling} ($d_s=2$). 

In conclusion, we have shown how to reconstruct the super-area law of the EE through the asymptotic behavior of the EC data distributed on the EF pattern, especially providing insight into the origin of the logarithmic correction for area law in the EE formula. In comparison, as discussed in \textbf{Appendix}~\ref{app_suptran_f}, the logarithmic correction for the case of translationally invariant Euclidean lattices arises from the application of the Widom conjecture of the asymptotic behavior of Toeplitz matrices. 

\section{The super-area law of free-fermion systems on translationally invariant lattice}\label{app_suptran_f}
In this section, we study the scaling behavior of the EE in free-fermion systems with translation invariance. The correlation matrix $C^{A}$, defined in the main text, typically manifests as a Toeplitz matrix, making it challenging to discern its asymptotic spectrum. Consequently, even in translationally invariant lattices, analytically studying the scaling of EE poses significant difficulties. Ref.~\cite{leePositionmomentum2014} provides an alternative definition of the correlation matrix $C^{A}$ for the subsystem $A$. By employing a projection operator $\hat{R}=\sum_{x\in \Omega}\ket{x}\bra{x}$ onto the lattice set $\Omega$, $C^{A}$ can be succinctly defined as:
\begin{equation}
    C^{A} = \hat{R}\hat{P}\hat{R},
\end{equation}
where $\hat{P}$ is also a projection operator that projects onto the occupied states via $\hat{P} = \theta(-\mathcal{H}) = \sum_{k}\theta(-\varepsilon_{F})\ket{\psi_{k}}\bra{\psi_{k}}$. Here, $\ket{\psi_{k}}$ and $\varepsilon_{k}$ denote the eigenvectors and eigenvalues of the single-particle Hamiltonian matrix $\mathcal{H}$ with the momentum $k$, and $\theta$ represents the step function, indicating occupied single-particle states with energy $\varepsilon_{k}\leq\varepsilon_{F}$. Based on the expression of EE $S_{A}$ in free-fermion systems in Eq.(15) of the main text, we can rewrite EE as follows:
\begin{equation}
    \begin{split}
    S_{A}&=\text{Tr}(\rho_{A}\ln\rho_{A}) \\ 
    &=-\sum_i [\xi_i \ln \xi_i+(1-\xi_i)\ln(1-\xi_i)] \\
    &= \sum_{i} f(\xi_{i}) \\
    &= \text{Tr}\left[f(\hat{R}\hat{P}\hat{R})\right], 
    \end{split}
\end{equation}
where the function $f(t) = -t\ln t - (1-t)\ln(1-t)$. Ref.~\cite{gioevEntanglement2006} discusses that the sum of function $f$ acting on the spectrum of the Hermitian operator $\hat{R}\hat{P}\hat{R}$ equals the trace of the function $f$ acting on the Hermitian operator $\hat{R}\hat{P}\hat{R}$. Meanwhile, the kernel of the operator $\hat{R}\hat{P}\hat{R}$ is given by
\begin{equation}
    \bra{\psi_{k}}\hat{R}\hat{P}\hat{R}\ket{\psi_{k^{'}}}=\chi_{\Gamma}(k)\chi_{\Gamma}(k^{'})\frac{V}{(2\pi)^{d}}\int_{\Omega}e^{i(k-k^{'})x}dx,
    \label{eq_kernel}
\end{equation}
where $\chi_{\Gamma}(k)$ is defined for $\Gamma$ as $\chi_{\Gamma}(k)=1$ if $k\in \Gamma$ and $\chi_{\Gamma}(k)=0$ otherwise. In a $d$-dimensional system with a $(d-1)$-dimensional Fermi surface, the Fermi sea region of the system is represented as $\Gamma=\{k| \varepsilon_{k}\leq \varepsilon_{F} \}$ in momentum space. Ref.~\cite{gioevEntanglement2006} discusses that when $L\rightarrow \infty$, the scaling of the system is determined by the Widom conjecture.

Next, we employ the Widom conjecture to derive the super-area law for $d$-dimensional systems with $(d-1)$-dimensional Fermi surfaces. In Ref.~\cite{gioevEntanglement2006}, considering the operator $\hat{R}\hat{P}\hat{R}$ defined in two sets $\Omega$ and $\Gamma$, with the kernel in Eq.~\eqref{eq_kernel}, and a general class of functions $f$, the asymptotic formula of the trace for $\hat{R}\hat{P}\hat{R}$ is derived as follows:
\begin{equation}
    \begin{aligned}
    &\text{Tr}f(\hat{R}\hat{P}\hat{R}) = \left(\frac{L_{A}}{2\pi}\right)^{d}f(1) \int_{\Omega}\int_{\Gamma}dx dk \\
    &\quad+ \left(\frac{L_{A}}{2\pi}\right)^{d-1}\frac{\ln 2\ln L_{A}}{4\pi^{2}}U(f)\int_{\partial \Omega } \int_{\partial \Gamma }|\bm{n}_{x}\cdot \bm{n}_{k}|d S_{x} d S_{k} \\
    &\quad+ o(L_{A}^{d-1}\ln L_{A}),
    \end{aligned}
    \label{eq_wid_conj}
\end{equation}
where $\partial \Gamma$ and $\partial \Omega$ represent the boundaries of the Fermi sea and the subsystem $A$ respectively. $\bm{n}_{x}$ and $\bm{n}_{k}$ denote unit normal vectors to $\partial \Gamma$ and $\partial \Omega$ respectively, and $U(f) = \int^{1}_{0}\frac{f(t)-tf(t)}{t(1-t)}dt$. Utilizing this formula, we readily obtain the scaling of EE in $d$-dimensional systems with $(d-1)$-dimensional Fermi surfaces. From Eq.~\eqref{eq_wid_conj}, the first term vanishes due to $\lim_{x\rightarrow 1^{-}}f(x)=0$. The second term of Eq.~\eqref{eq_wid_conj} determines the scaling of EE and is proportional to $L_{A}^{d-1}\log L_{A}$. It is essential to note that this formula holds true only when $\hat{R}\hat{P}\hat{R}$ is Hermitian. In conclusion, by leveraging the Widom conjecture, we establish the logarithmic correction for area law of the scaling of EE on translationally invariant lattice.

\end{document}